\documentclass[final,5p,twocolumn]{elsarticle}

\usepackage{amssymb}
\usepackage{amsmath}
\usepackage{graphicx}
\usepackage{xcolor}
\usepackage{array}
\usepackage{caption}
\usepackage{placeins}
\usepackage[symbol]{footmisc}
\usepackage{wrapfig}

\journal{Photoacoustics}

\begin{document}

\begin{frontmatter}

\title{Conoscopic interferometry for optimal acoustic pulse detection in ultrafast acoustics}

\author[inst1]{Martin Robin\corref{cor1}}
\ead{martin.robin@univ-lyon1.fr}
\cortext[cor1]{Corresponding author. Present address: Université Claude Bernard Lyon 1, CNRS, Institut Lumière Matière, Villeurbanne, F-69622, France}

\author[inst1]{Ruben Guis}

\author[inst2]{Mustafa Umit Arabul}
\author[inst2]{Zili Zhou}
\author[inst2]{Nitesh Pandey}

\author[inst1]{Gerard J. Verbiest\corref{cor2}}
\ead{G.J.Verbiest@tudelft.nl}
\cortext[cor2]{Corresponding author}

\address[inst1]{Department of Precision and Microsystems Engineering, Delft University of Technology, Mekelweg 2, Delft, 2628CD, Netherlands}
\address[inst2]{ASML Netherlands B.V., De Run 6501, Veldhoven, 5504DR, Netherlands}

\begin{abstract}
Conoscopic interferometry is a promising detection technique for ultrafast acoustics. By focusing a probe beam through a birefringent crystal before passing it through a polarizer, conoscopic interferences sculpt the spatial profile of the beam. The use of these patterns for acoustic wave detection revealed a higher detection sensitivity over existing techniques, such as reflectometry and beam distortion detection. However, the physical origin of the increased sensitivity is unknown. In this work, we present a model, describing the sensitivity behaviour of conoscopic interferometry with respect to the quarter-wave plate orientation and the diaphragm aperture, which is validated experimentally. Using the model, we optimize the detection sensitivity of conoscopic interferometry. We obtain a maximal sensitivity of detection when placing the diaphragm edge on the dark fringes of the conoscopic interference patterns. In the configurations studied in this work, conoscopic interferometry can be 8x more sensitive to acoustic waves than beam distortion detection.
\end{abstract}


\end{frontmatter}

\section{Introduction}
Photoacoustics uses pulsed lasers to excite high-frequency acoustic waves ranging from hundreds of kHz to hundreds of GHz \cite{White1963}\cite{Thomsen1984} for non-destructive testing \cite{Scruby1989}, material characterization \cite{Sermeus2012}, and for medical imaging and diagnosis \cite{LaCaveraIII2021}. Usually, a nanosecond (ns) \cite{Faese2013} to femtosecond (fs) \cite{Devos2004} pulsed laser – the pump - generates bulk, guided, or surface acoustic waves in a sample of interest \cite{Royer2000}. The detection of the same acoustic waves with a second laser beam – the probe - enables non-contact measurements on samples with complex geometries, in tough environmental conditions, and without contaminating their surface \cite{Scruby1989}.

The most common implementation for acoustic wave detection with lasers is reflectometry. The strain associated with the acoustic waves changes the refractive index of the material through the photoelastic effect \cite{Matsuda2015}. Hence, the power of the probe beam reflected from the material surface has a component directly proportional to the elastic strain. The resulting relative variation in laser power is usually in the range of $10^{-6}- 10^{-4}$ \cite{Thomsen1984} \cite{Chigarev2006} \cite{Liu2018}. The photoelastic constants of the material at the probe laser wavelength set the detection sensitivity.

The strong dependence of the photoacoustic signal on the photoelastic constants limits the applicability of reflectometry and thus inspired the development of Beam Distortion Detection (BDD) \cite{Chigarev2006} \cite{Higuet2011} and Conoscopic Interferometry (CI) \cite{Liu2018}. In BDD, the Gaussian spatial profile of the acoustic wave incident on the sample surface causes slight fluctuations in the divergence angle of the reflected probe beam.
This results in diameter variations of the reflected probe beam that are proportional to the acoustic displacement, hence variations in power density. By masking a part of the probe beam with a diaphragm, the power measured with a photodetector becomes proportional to the displacement of the sample surface. This technique has the advantage of a detection sensitivity \emph{independent} of the properties of the sample material: BDD does allow the detection of acoustic waves in materials with very low photoelastic constants. In this case, Chigarev {\it et al.} reported a clear improvement of the Signal-to-Noise Ratio (SNR) with respect to reflectometry \cite{Chigarev2006}. In general, the measured signal is a sum of the BDD and the reflectometry signal. In materials with high photoelastic coefficients, BDD and reflectometry signals are therefore difficult to distinguish from each other \cite{Chigarev2006} \cite{Chigarev2007}.

CI makes use of Conoscopic Interference Patterns (CIPs), which are well-known for the characterization of birefringent crystals \cite{Liu2018} \cite{Ayras1999} \cite{Wang2012}. By focusing the probe beam with a given polarization through a birefringent crystal and then collimating it before passing through a polarizer, one can obtain a succession of bright (isochromates) and dark (isogyres) fringes. The fringes form a pattern characteristic of the birefringence properties of the crystal and the input and output polarizations. Liu {\it et al.} \cite{Liu2018} implemented this phenomenon in BDD by adding a birefringent crystal (sapphire plate) between the objective and the sample and by using a Polarizing Beam Splitter (PBS) as a polarizer to change the spatial profile of the probe beam to a CIP. The resulting CIP is controlled by rotating a quarter-wave plate placed between the PBS and the objective. Liu {\it et al.} observed surprisingly high SNR for some CIPs, with respect to BDD and reflectometry in identical configurations on two different samples. However, to this day, an analytical model that predicts the sensitivity of acoustic wave detection by CI is still missing. This is of great interest in view of pushing the sensitivity higher to allow measurements of weak acoustic signals from thick structures, reflections from interface with low acoustic impedance mismatch, or materials with high acoustic damping.

In this paper, we present an analytical model for the CIPs and predict their sensitivity to acoustic waves. Using this model, we identified the key parameter to optimize the performance of CI: maxima of sensitivity is obtained by placing the diaphragm edges on the dark fringes (isogyres) of the CIPs. In the configurations considered in this work, we found a sensitivity up to 8 times higher than that of BDD. We experimentally validate the model on a 2.4 \textmu m thick silicon plate (Si) coated with $\sim 30$ nm aluminum (Al) indicating that the model correctly predicts the sensitivity of CI to acoustic waves.

\begin{figure}[h]
\includegraphics[scale=1]{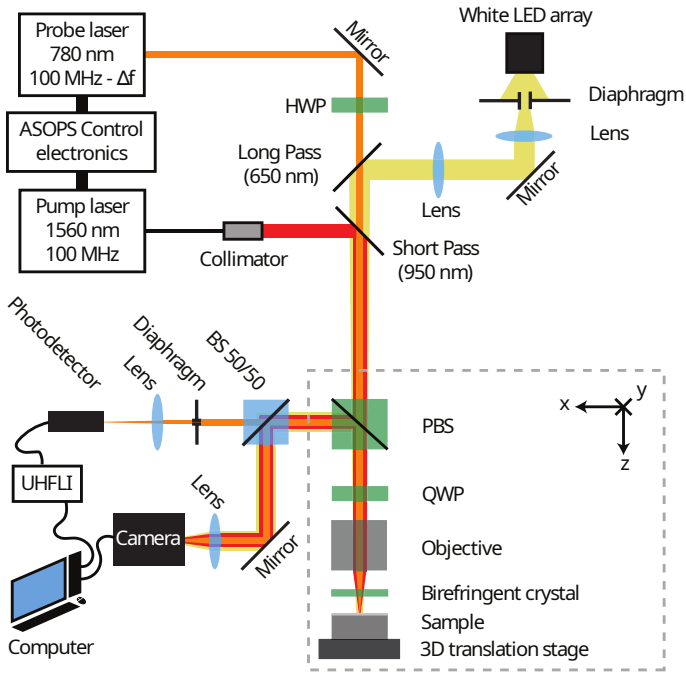}
\centering
\caption{Picosecond ultrasonics ASOPS setup with Conoscopic Interferometry detection.
HWP: Half-Wave Plate, QWP: Quarter-Wave Plate, (P)BS: (Polarizing) Beam Splitter, LP/SP: Long Pass/Short Pass dichroic mirrors (cut-off wavelength).
The UHFLI from Z\"{u}rich Instruments records and analyzes the photodetector signal before sending it to the computer. A white LED array illuminates the sample. We use a camera for aligning the lasers with respect to each other and the sample.}
\label{fig:fig1}
\end{figure}

\section{Materials and methods}

The experimental setup, shown in Figure \ref{fig:fig1}, contains an ASynchronous OPtical Sampling (ASOPS) system \cite{Bartels2007} consisting of two synchronized Erbium lasers from Menlo Systems with a pulse duration of around 100 fs. The pump pulses locally heat the sample, which results in an extremely short temperature increase and to the thermomechanical generation of a longitudinal acoustic pulse \cite{Matsuda2015} in a bandwidth of a few tens of GHz ($\sim 10-100$ GHz). The probe pulses allow us to measure the acoustic reflections arriving back at the surface of the sample. The pump laser has a wavelength of 1560 nm, a repetition rate of 100 MHz, and an average output power of around 100 mW. The probe laser has a wavelength of 780 nm, an average output power of $\sim 500$ \textmu W, and a $\sim 10$ kHz lower repetition rate than the pump laser. This offset in repetition rate allows the reconstruction of the 10 ns time window between two pump pulses within 100 \textmu s. The time window is thus probed with $10^4$ discrete time samples and consequently offers a temporal resolution of 1 ps. In this section, we describe the experimental setup by introducing successively the paths of the pump beam, probe beam and the illumination of the sample as well as the data acquisition and measurement methodologies. 

\subsection{Pump beam path}

The pump beam fiber output is first collimated by a collimator and reflected at 90° by a Short-Pass (SP) 950 nm dichroic mirror to make a common path with the probe beam. The P-polarized component is transmitted by a polarized beam splitter (PBS) and then crosses a Quarter-Wave Plate (QWP) before it is focused on the sample through a sapphire plate by an objective. The near-infrared long working distance Plan-Apochromat objective from Mitutoyo has a magnification of 20 and a wavelength correction from visible range to 1800 nm. A part of the pump beam reflected off the sample is redirected towards a camera using a Beam Splitter (BS) for aligning the pump and probe beam. The radius of the pump beam, defined by the Half Width Half Maximum (HWHM) of the intensity, is estimated as $r_{pu} \approx 2$ mm directly after the collimator and $a_{pu} \approx 2$ \textmu m on a sample in focus.

\subsection{Probe beam path}

The probe beam is free-space and first passes through a Half-Wave Plate (HWP) to make it P-polarized. This maximizes the power transmitted by the PBS. Before crossing the PBS, the probe beam travels through two dichroic mirrors; a Long-Pass (LP) with a cut-off wavelength of 650 nm and the SP with a cut-off wavelength of 950 nm. We place these dichroic mirrors before the PBS to avoid any shift in polarization of the probe beam after the PBS as this would affect the CIPs. After crossing the PBS, we place a QWP to controllably rotate the probe beam polarization.

The objective focuses the probe beam on the sample through a 1 mm thick C-axis cut (0001) birefringent sapphire plate, which modifies the beam polarization and gives it a spatial dependence. After reflection of the probe beam by the sample, it passes again through the sapphire plate, the objective, and the QWP. Now, the PBS acts as a polarizer and reflects the S-polarized component only towards the detection arm of the setup. 

In the detection arm, a BS splits the probe beam into two beams of equal power. One of these beams is focused on a camera, to visualize the CIPs and to align the pump and probe beam on the sample. The other beam is truncated by an iris diaphragm, of which the aperture diameter can be set between 0.4 mm and 8 mm. This diaphragm is used to detect the acoustic waves in a BDD or CI configuration. After the diaphragm, the probe beam is focused on a photodetector to ensure a spot size smaller than the photosensitive area and thereby avoid additional truncation of the beam. The probe beam radius (HWHM) is estimated as $r_{pr} \approx 0.4$ mm directly at the laser output and $a_{pr} \approx 3$ \textmu m on a sample in focus.

\subsection{Sample illumination}

A white LED array illuminates the sample to localize the pump and probe beam spots with respect to the sample. The white light of the LED is first collimated to a beam by the use of a lens and a diaphragm. This allows us to control the white beam’s diameter and power by adjusting its aperture. A lens with a long focal length then focuses the white beam to avoid loss of power by truncation on the aperture of the other optical components in the setup.
Before reaching the sample, the white beam is first reflected with an angle of 90° by the LP 650 nm dichroic mirror and then crosses the SP dichroic mirror, the PBS, the QWP, the objective, and the sapphire plate. The white beam is then reflected by the surface of the sample and crosses the sapphire plate, the objective, the QWP, and the PBS before moving into the detection arm of the setup. Part of the white beam is reflected by the BS and is focused on the camera to visualize the position of the pump and probe spots with respect to the sample.

\begin{figure}[h]
\includegraphics[scale=1]{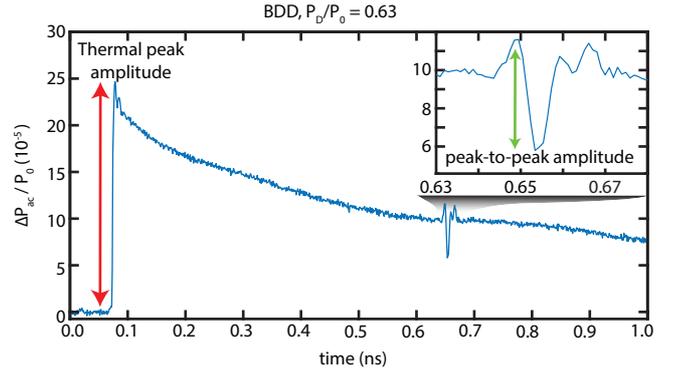}
\centering
\caption{Typical signal ($\Delta P_{ac}/P_0$ vs. time) measured in BDD for a diaphragm aperture of $P_D/P_0$ = 0.63. Inset: zoom on the first acoustic echo. We use the amplitude of the thermal peak and the peak-to-peak amplitude of the first acoustic echo to quantify the measurements sensitivity.}
\label{fig:fig2}
\end{figure}

\subsection{Data acquisition}

We detect the probe beam pulses using a Si amplified photodetector from Menlo System (FPD510-FS-VIS) that is sensitive in a wavelength range from 400 nm to 1000 nm and has a bandwidth of 250 MHz. The photosensitive area of the photodetector has a diameter of 0.4 mm.
The signal coming from the photodetector is processed by a lock-in amplifier (Ultra High Frequency Lock-In amplifier from Zürich Instruments, 600 MHz bandwidth) with the Boxcar + Periodic Waveform Analyzer function \cite{ZI}. This allows the accurate reconstruction of the individual probe pulses. By using a trigger signal from the ASOPS system at a frequency equal to the difference in the repetition rate between both lasers ($\sim 10$ kHz), we probe the full-time delay window from 0 to 10 ns. The measured signals correspond to a variation in the probe pulse power induced by the response of the sample. We normalize these signals by dividing them by the probe power incident on the photodetector when the diaphragm is fully open.

\subsection{Measurement methodology for BDD and CI}
\label{sec2.5}

To investigate the influence of the diaphragm aperture in the probe beam path on the detection sensitivity of CI, we study two experimental configurations:
\begin{itemize}
    \item BDD configuration: without the sapphire plate present in the setup (Figure \ref{fig:fig1}), to validate the experimental methodology in the well-known BDD case.
    \item CI configuration: using a sapphire plate with a thickness $h$=1 mm and three different orientations of the QWP's fast axis orientation, $\theta_{1/4}$ = 0°, 25°, 45°.
\end{itemize}
We perform measurements on a 2.4 \textmu m thick Si sample coated with $\sim 30$ nm of Al (Atomic Force Microscopy probe, model CONTR from NanoWorld).
By measuring the full probe power $P_0$ with the diaphragm fully open before each measurement, we ensure that $P_0$ is the same for all the measurements. We determine the power ratio $P_D/P_0$ between the power after ($P_D$) and before the diaphragm by measuring the power incident on the photodetector after partly closing the diaphragm.
Depending on the configuration, between 7 and 9 diaphragm aperture diameters are used, ranging from $P_D/P_0 = 0$ to 1.

The pump and probe lasers are both focused on the free surface of the Al film. Figure~\ref{fig:fig2} shows a typical measurement of the relative variation of probe power incident on the photodetector $\Delta P_{ac}/P_0$ induced by the response of the sample to the pump pulse. The thermal response starts at $\sim 0.08$ ns, consisting in a peak due to the very fast temperature increase and then an exponential decay due to cooling.
The acoustic reflection coefficient between the aluminium and silicon is very low (\textless 1\% of the acoustic energy), which makes the amplitude of the acoustic reflection at the Al/Si interface very weak. Therefore, we observe the first clear acoustic reflection (longitudinal wave) from the backside of the sample, arriving at $\sim0.65$ ns. The time delay between the thermal peak and the acoustic echo (0.57 ns) corresponds to a Si thickness of 2.4 \textmu m, which is within the range specified by Nanoworld. In each acquisition, we extract the amplitude of the thermal peak as well as the peak-to-peak amplitude of the first acoustic echo to quantify the sensitivity of detection.

\subsection{Measurement methodology for reflectometry}

For a good comparison between reflectometry and BDD/CI measurements, we pay particular attention to the distinct contributions of reflectometry and BDD/CI components to the experimental signals. The aluminium in the sample offers an interband transition around 780 nm \cite{Devos2003} \cite{Humbert1977} resulting in high photoelastic constants at the probe wavelength and thus in a high reflectometry component. The measured signals are therefore a sum of the reflectometry and the BDD/CI contributions, as explained by Chigarev {\it et al.} \cite{Chigarev2006}.

To extract the BDD contribution, we compare the signals for BDD with similar measurements performed only in reflectometry. For these reflectometry measurements, we completely open the diaphragm and decrease the power at the output of the probe laser until we have the same incident power on the photodetector as in the corresponding BDD measurement. This emulates the loss of power induced by the diaphragm. The measurements are normalized in the same way as for the BDD measurements. Since reflectometry is based on the variation of the local refractive index by the acoustic strain \cite{Matsuda2015}, its sensitivity is directly proportional to the probe power incident on the photodetector.

Similar to BDD and CI measurements, we use the thermal peak amplitude and the peak-to-peak amplitude of the first acoustic reflection inside the sample to characterize the sensitivity of reflectometry (see Figure \ref{fig:fig2}).

\section{Theory}
\label{sec3}

The analytical model we present here combines the influence of the diaphragm aperture on the acoustic wave detection sensitivity in BDD \cite{Chigarev2006} with the Jones calculus formulation for the CIPs \cite{Liu2018} \cite{Ayras1999} \cite{Wang2012}. BDD is thus a particular case of the model, where the probe beam is spatially Gaussian. The full model can be applied to any kind of beam shape.

Without loss of generality, we assume a probe beam that is spatially uniform and purely S-polarized. In the experimental setup, after crossing the HWP, the PBS, the QWP, the sapphire plate and reflection by the sample, the electric field $\vec{E}$ of the beam is as follows:

\begin{equation}
    \vec{E}(x,y) = M_R W_{S}(x,y) W_{1/4} P_{t} W_{W1/2} 
    \begin{pmatrix}
        0\\
        1
    \end{pmatrix},
    \label{eq1}
\end{equation}

\noindent
wherein $\begin{pmatrix} 0, 1 \end{pmatrix}^T$, represents the S-polarized beam at the output of the laser. $W_{n}$ is the Jones calculus formulation of the wave plates ($n = 1/4, 1/2$, indicating if it is a QWP or a HWP and $n = S$ indicating the sapphire plate), and  $P_{t} = \begin{pmatrix} 1 & 0 \\ 0 & 0 \end{pmatrix}$ is the one of the PBS in transmission for the P-polarized component. The matrix $M_R =\begin{pmatrix} -1 & 0 \\ 0 & 1 \end{pmatrix}$ models the reflection of the probe beam on the sample, which simply acts as a mirror \cite{Liu2018}. $W_{n}$ is defined as follows:

\begin{equation}
    W_{n} = R(\theta_n)^T
    \begin{pmatrix}
        1 & 0 \\
        0 & e^{-j\delta_n}
    \end{pmatrix}
    R(\theta_n),
    \label{eq2}
\end{equation}

\noindent
wherein $\theta_n$ is the angle of the QWP's or HWP's fast axis with respect to the x axis, $R(\theta_n)$ the corresponding rotation matrix (and $R(\theta_n)^T$ its transpose), and $\delta_n$ the phase shift of the wave plate. The phase shift $\delta_n$ equals $\pi$ for the HWP and $\pi/2$ for the QWP. The angle of orientation $\theta_{1/4}$ of the QWP controls the CIPs \cite{Liu2018}.
The sapphire plate acts as a wave plate due to its birefringent properties \cite{Liu2018} \cite{Wang2012}  and is represented by the matrix $W_{S}$. As the probe beam is focused through the sapphire plate, the angle $\theta_S$ and phase shift $\delta_S$ depend on the location in the $(x,y)$ plane. Therefore, unlike for the HWP and the QWP, $\theta_S$ and $\delta_S$ in $W_{S}(x,y)$ are position dependent and are expressed as follows \cite{Liu2018}:

\begin{equation}
    \theta_S(x,y) = \tan^{-1}(y/x),
    \label{eq3}
\end{equation}

\begin{equation}
    \delta_S(x,y) = \frac{2\pi}{\lambda}h(n_e-n_o)\sin^2\left(\tan^{-1}\left(\frac{\sqrt{x^2+y^2}}{f_{obj}}\right)\right).
    \label{eq4}
\end{equation}

\noindent
Here, $\lambda = 780$ nm is the wavelength of the probe laser, $h$ is the thickness of the sapphire plate, $n_e = 1.760$ and $n_o = 1.768$ are the extraordinary and the ordinary refractive indices, respectively, and $f_{obj} = 20$ mm is the focal length of the objective.

Since the pump beam is spatially Gaussian, the displacement induced by the acoustic pulse when it reaches the sample’s surface is Gaussian as well. This produces a slight variation in the reflected probe beam divergence angle, as explained in \cite{Chigarev2006}. The relative variation $\xi$ on the objective plane of the reflected probe beam radius $r_{pr}^{'}$ with respect to the incident beam radius $r_{pr}$ (considering $r_{pr}^{'} = (1+\xi) r_{pr}$) is \cite{Chigarev2006}:

\begin{equation}
    \xi = 2 \frac{2\pi z_0 a_{pr}^2}{\lambda z_r a_{pu}^2} A_0,
    \label{eq5}
\end{equation}

\noindent
where $z_0$ is the distance between the sample position and the probe beam focus position, $z_r \approx 35$ \textmu m is the Rayleigh length of the probe beam, and $A_0$ is the displacement amplitude of the sample’s surface due to the acoustic wave.

Before reaching back the objective, the probe beam again crosses the sapphire plate. The variation in the reflected probe beam divergence angle caused by the acoustic wave induces a shift in coordinate on the sapphire plate with respect to the probe beam which was initially incident on the sample. Furthermore, due to the reverse propagation direction of the reflected light, the orientation of the fast axis with respect to the beam is mirrored with respect to the beam incident on the sample. Therefore, we now use $\theta_{S,r} = \pi - \theta_S$ and $\delta_{S,r}(x,y) = \delta_{S}((1+\xi)x,(1+\xi)y)$ in $W_{S}(x,y)$ to obtain $W_{S,r}(x,y)$.
After being collimated by the objective, the probe beam crosses the QWP with a reverse propagation direction (i.e., $\theta_{r,1/4} = \pi - \theta_{1/4}$ in $W_{1/4,r}$) and the PBS. The PBS now reflects the S-polarized component of the beam, $P_{r} =  \begin{pmatrix} 0 & 0 \\ 0 & 1 \end{pmatrix}$. All combined, this results in the following expression for the electric field $\vec{E}$ arriving at the photodetector:

\begin{eqnarray}
    \vec{E}(x,y) = &P_{r}W_{1/4,r}W_{S,r}(x,y)M_R \times\nonumber\\
     & W_{S}(x,y) W_{1/4}P_{t}W_{1/2}
     \begin{pmatrix}
        0\\
        1
    \end{pmatrix}.
    \label{eq6}
\end{eqnarray}

From the electric field $\vec{E}$, we obtain the probe beam intensity $I(x,y)$ incident on the photodetector: $I(x,y) = c \epsilon_0 |\vec{E}(x,y)|^2 /2$, where $c = 3*10^8 m.s^{-1}$ is the speed of light, and $\epsilon_0 = 8.85*10^{-12} F.m^{-1}$ is the vacuum permittivity. The spatial dependence of the intensity directly gives us the CIPs induced in the probe beam. In reality, the probe beam is spatially Gaussian at the output of the laser. Therefore, the beam intensity as seen by the photodetector becomes in polar coordinates $(r,\phi)$:

\begin{equation}
    I_{G}(r,\phi) = I(r,\phi) e^{-r^2/(r_{pr} (1+\xi))^2} /(\pi (r_{pr} (1+\xi))^2 ).
    \label{eq7}
\end{equation}

Finally, we take into account the influence of the diaphragm, for which we assume a circular aperture perfectly aligned with the center of the beam. The relative variation of probe power $\Delta P_{ac}/P_0$ incident on the photodetector is given by:

\begin{equation}
    \frac{\Delta P_{ac}}{P_0} = \frac{\int_0^{r_D}\int_0^{2\pi}(I_{G}(r,\phi)-I_{G,0}(r,\phi))r dr d\phi}{\int_0^{\infty}\int_0^{2\pi}I_{G,0}(r,\phi) r dr d\phi},
    \label{eq8}
\end{equation}

\noindent
where $r_D$ is the radius of the diaphragm aperture, $\infty$ represents the radius of the diaphragm aperture when it is completely open, and $I_{G,0}(r,\phi)$ is the probe beam intensity incident on the photodetector when the sample is not excited by the pump beam. Equation \ref{eq8} directly gives the relative variation in probe power induced by the acoustic waves in presence of a CIP and a diaphragm.

By removing the Gaussian profile from Equation \ref{eq7} and the acoustic wave contribution ($\xi = 0$), we find back the CIPs as presented by Liu {\it et al.} in \cite{Liu2018}. By assuming a non-birefringent crystal ($n_e=n_o$ and thus $\delta_S= \delta_{S,r} = 0$), Equation \ref{eq8} reduces to the BDD signal as derived by Chigarev {\it et al.} \cite{Chigarev2006}. BDD is thus a particular case of Equation \ref{eq8} when the crystal used is not birefringent.

The theory resulting in Equation \ref{eq8} highlights the main parameters influencing the sensitivity of CI to acoustic waves:
\begin{itemize}
    \item The QWP orientation $\theta_{1/4}$.
    \item The refractive indexes, $n_e$ and $n_o$ of the birefringent crystal.
    \item The thickness $h$ of the birefringent crystal.
    \item The angle of the focused probe beam w.r.t. the birefringent crystal set by the focal length $f_{obj}$ and thus the probe beam radius on the objective $r_{pr}$.
    \item The position of the sample $z_0$ with respect to the probe beam focus.
    \item The ratio between the probe and pump spot radii on the sample, $(a_{pr}/a_{pu})^2$.
    \item The diaphragm aperture $r_D$ with respect to the beam radius $r_{pr}$.
\end{itemize}
	
For the calculations below, we set $z_0 = -0.5$ \textmu m and $A_0 = 0.1$ nm. The precision of the translation stage used to adjust the position of the sample provides a resolution of 0.5 \textmu m for $z_0$. Although the pump laser characteristics and the sample material and geometry determine $A_0$, its value is typically of the order of several tenths of pm \cite{Chigarev2006}. Due to the negative value of $z_0$, $\xi$ is thus negative in our calculations below.

\begin{figure*}[h]
\includegraphics[scale=1]{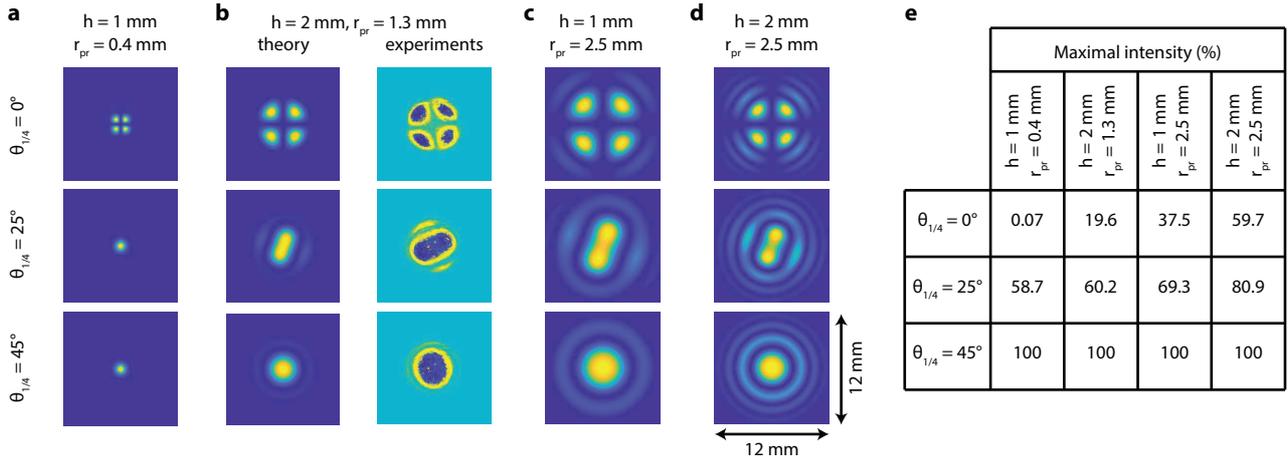}
\centering
\caption{Conoscopic interference patterns for $\theta_{1/4}$ = 0°, 25°, 45° and for different values of $r_{pr}$ and $h$: {\bf a} $h = 1$ mm and $r_{pr} = 0.4$ mm, {\bf b} $h = 2$ mm and $r_{pr} = 1.3$ mm, {\bf c} $h = 1$ mm and $r_{pr} = 2.5$ mm, and {\bf d} $h = 2$ mm and $r_{pr} = 2.5$ mm. Panel {\bf b} shows both the theoretical and measured conoscopic interference patterns. All shown conoscopic interference patterns have a physical size of 12x12 mm$^2$. {\bf e} Table with the maximum intensity value $I_{G,0}$ ($\%$) for each calculated pattern relative to an input intensity of $7171 W/m^2$.}
\label{fig:fig3}
\end{figure*}

\section{Results and discussion}

The results and discussion section is organized as follows. We present in Section~\ref{41} the calculated CIPs and validate them with the experiment. In Section~\ref{42} we show good agreement between both the theoretical and experimental sensitivity of CI and BDD to the acoustic waves. Finally, in Section~\ref{43}, we elucidate the dependence of the sensitivity of CI to the probe beam radius $r_{pr}$, the QWP orientation $\theta_{1/4}$, and diaphragm aperture size $P_D/P_0$, in order to optimize the sensitivity.

\subsection{Conoscopic interference patterns}
\label{41}

Figure~\ref{fig:fig3} shows CIPs for several probe beam radii and thicknesses of the sapphire plate. For each configuration, we show three QWP orientations corresponding to $\theta_{1/4}$ = 0°, 25° and 45°. The patterns calculated in the Figure \ref{fig:fig3}a correspond to the configuration studied experimentally in the Section~\ref{42}, with $r_{pr} = 0.4$ mm and $h = 1$ mm. In this configuration, the phase shift $\delta_S(x,y)$ induced by the sapphire plate (Equation \ref{eq4}) is only -1.5° for light leaving the objective at a distance $r_{pr}$ from the optical axis. The patterns observed for $\theta_{1/4}$ = 25°, 45° are very close to a spatial profile of a purely Gaussian beam. For $\theta_{1/4}$ = 0°, the pattern is different, showing bright (isochromates) and dark fringes (isogyres). For $\theta_{1/4} = $ 25°, 45°, the polarization is elliptical and circular, respectively, whereas the beam is purely P-polarized when $\theta_{1/4} = $ 0°. Since the PBS reflects only the S-polarized component towards the detection arm of the setup, and since the sapphire plate does not induce a phase shift at the center of the probe beam, this results in an isogyre. The influence of a weak phase shift $\delta_S$ between the P and S-polarized components of the beam, is therefore only clearly visible when $\theta_{1/4} =$ 0°. Due to this, the CIPs for $\theta_{1/4}$ = 25°, 45° also have an intensity $1000\times$ higher than for $\theta_{1/4}$ = 0° (see Figure~\ref{fig:fig3}e).

To enable the experimental observation of the CIPs, we use a sapphire plate with $h = 2$ mm and a beam expander directly at the probe laser output to increase the diameter to $r_{pr} \approx 1.3$ mm. Consequently, the maximum value $\delta_S(x,y)$ at a distance $r_{pr}$ from the center of the beam in this configuration increases to -31°, which induces a more significant difference between the CIPs and a Gaussian profile. The calculated and measured CIPs are presented in Figure \ref{fig:fig3}b. Note that the beam expander reduces the ratio between $a_{pr}$ and $a_{pu}$, which decreases the sensitivity of CI and BDD (Equation \ref{eq5}, \cite{Chigarev2006}) and therefore we do not consider this configuration in Section \ref{42}. For $\theta_{1/4}$ = 0°, the pattern is similar to that in Figure \ref{fig:fig3}a, but the CIPs for $\theta_{1/4}$ = 25° and $\theta_{1/4}$ = 45° are different. The $\theta_{1/4}$ = 25° loses its circular symmetry and both the $\theta_{1/4}$ = 25° and $\theta_{1/4}$ = 45° CIPs contain fringes around a central maximum in intensity. The intensity of the different patterns is also now of the same order of magnitude ($\sim 1000 W/m^2$, see Figure~\ref{fig:fig3}e). Due to the initially spatially Gaussian profile of the probe beam, the fringes of the CIPs have a lower intensity than their centers (see Equation~\ref{eq7}). We observe the same features and patterns experimentally which validates the model presented in Section~\ref{sec3}. We attribute the slight rotation between the patterns obtained theoretically and experimentally to the unknown reference coordinate for the polarization of the probe beam in the experiment.

To gain more insight into the parameters determining the CIPs, we plot them for different combinations of $r_{pr}$ and $h$ in Figures~\ref{fig:fig3}c and \ref{fig:fig3}d.
When $r_{pr}$ increases from $0.4$~mm to $2.5$~mm (Figures \ref{fig:fig3}a to \ref{fig:fig3}d), we observe the appearance of more bright (isochromates) and dark (isogyres) fringes around the central shape.
The appearance of more fringes is due to the increased convergence angle ($\tan^{\rm -1}(r_{pr}/f_{obj})$ in Equation \ref{eq4}) of the light passing through the sapphire plate. As a consequence, $\delta_S$ increases resulting in stronger conoscopic interferences and thus in the appearance of more bright and dark fringes in the pattern. 
For the same reason, also more bright and dark fringes appear in the CIP when increasing $h$ from 1 to 2 mm (Figures \ref{fig:fig3}c to \ref{fig:fig3}d).
Also, the CIP for $\theta_{1/4}$ = 25° in Figures~\ref{fig:fig3}c and \ref{fig:fig3}d clearly differs from that for $\theta_{1/4}$ = 45°.
This is due to the large $r_{pr}$ which ensures significant intensity in the bright fringes. 
As a result, all different CIPs in Figures~\ref{fig:fig3}c and \ref{fig:fig3}d show similar intensities.

\subsection{Sensitivity to sample deformations}
\label{42}

\begin{figure}[t]
\includegraphics[scale=1]{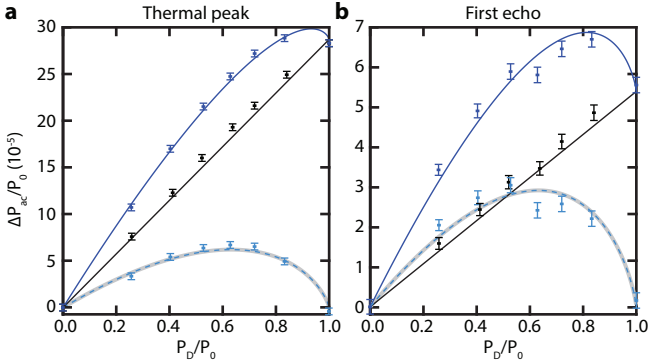}
\centering
\caption{Measurements of the detection sensitivity of BDD and reflectometry and associated fits: {\bf a} thermal peak amplitude and {\bf b} peak-to-peak amplitude of the first acoustic echo. The blue data points represent the BDD measurements and the corresponding continuous blue lines the fit to Equation~\ref{eq8}. Similarly, the black data points and lines show the reflectometry measurements and fit. The light blue data points and light dashed blue lines show the BDD measurement and fit from which the reflectometry component has been subtracted. The grey lines show the sensitivity obtained using the model of Chigarev et al. \cite{Chigarev2006} scaled to match the amplitude of the dashed light blue curve.}
\label{fig:fig4}
\end{figure}

\begin{table}[h]
\includegraphics[scale=1]{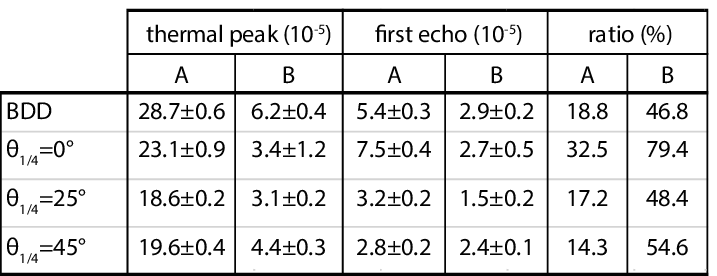}
\centering
    \captionof{table}{Values of the fit parameters, $A$ (reflectometry sensitivity) and $B$ (BDD/CI sensitivity) and their associated error, for the different experimental configurations studied. The last two colomns show the ratio (\%) between the $A$($B$) fit parameters of the first acoustic echo and the thermal peak.}
    \label{tab:TableA1}
\end{table}

To further validate the model presented in Section \ref{sec3}, we now focus on the measurements obtained in the BDD configuration and compare it to the sensitivity profile presented by Chigarev {\it et al.} \cite{Chigarev2006}. In order to identify the BDD contribution to the total signal, we compare the measurements presented in Figures~\ref{fig:fig4}a and \ref{fig:fig4}b for BDD (in blue) with similar measurements performed only in reflectometry (in black). Then, we fit the experimental data of the BDD configuration using the following function $f_{fit}$:

\begin{equation}
    f_{fit} (P_D/P_0) = A \cdot P_D/P_0 + B \cdot f_{th} (P_D/P_0),
    \label{eq9}
\end{equation}

\noindent
where $f_{th}$ is the theoretical sensitivity function of BDD or CI calculated from Equation~\ref{eq8} ($\Delta P_{ac}/P_0$) normalized to its maximum value. The fit parameters $A$ and $B$ correspond to the amplitude of reflectometry and BDD, respectively. As reflectometry is directly proportional to the probe power, $A$ is simply multiplied with $P_D/P_0$.
The continuous blue line in Figures~\ref{fig:fig4}a and \ref{fig:fig4}b shows the best fit result. The continuous black line depicts the reflectometry part ($A \cdot P_D/P_0$) and the dashed blue line the BDD contribution ($B \cdot f_{th} (P_D/P_0)$). For comparison, we also plot the experimental data points from which we subtracted the reflectometry component ($A \cdot P_D/P_0$), and the calculation (gray line) from the model of Chigarev et al. \cite{Chigarev2006}. The values of the fit parameters $A$ and $B$ are listed in Table \ref{tab:TableA1} and are similar to values for reflectometry on aluminium \cite{Devos2003} and BDD \cite{Chigarev2006,Liu2018} reported in the literature. Figures \ref{fig:fig4}a and \ref{fig:fig4}b thus show that the model of Section~\ref{sec3} predicts the sensitivity of both the thermal peak amplitude and the peak-to-peak amplitude of the first acoustic echo.

\begin{figure}
\includegraphics[scale=1]{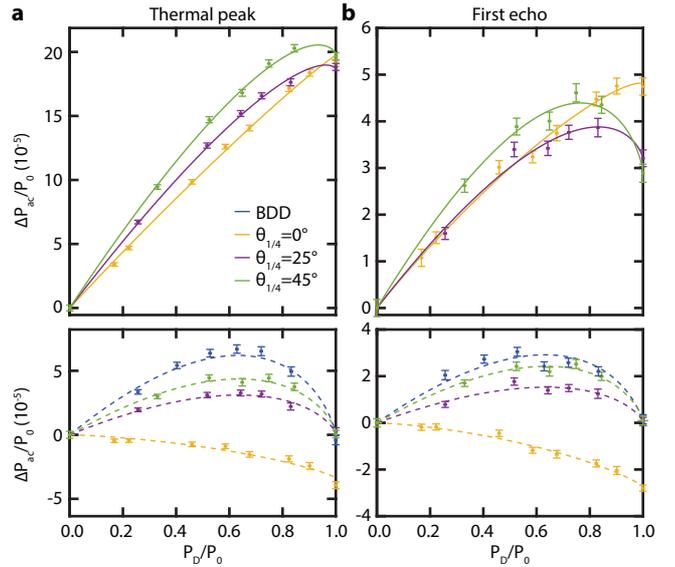}
\centering
\caption{Measurements of the sensitivity of detection of BDD and CI and associated fits: {\bf a} thermal peak, {\bf b} first acoustic echo. The top panels show the measured data points and fits to equation~\ref{eq8}. The lower panels show the measured data points and fits after subtracting the reflectometry component. For completeness, we show the BDD measurement of Figure~\ref{fig:fig4} in blue. The CI data are shown in yellow for $\theta_{1/4} =$ 0°, purple for $\theta_{1/4} =$ 25°, and green for $\theta_{1/4} =$ 45°.}
\label{fig:fig5}
\end{figure}

As the fitting procedure was validated for BDD, we now focus on the CI configuration with $h$ = 1 mm, $r_{pr}$ = 0.4 mm. In this configuration, we cannot measure the reflectometry contribution independently due to the sapphire plate. Therefore, we rely on the fitting procedure to separate the reflectometry contribution from the CI contribution.
Figure \ref{fig:fig5} shows the CI measurements before and after the subtraction of the reflectometry component. We compare three different orientations of the QWP (yellow: $\theta_{1/4}$=0°, purple: $\theta_{1/4}$=25°, green: $\theta_{1/4}$=45°) with the BDD measurements (blue). Note that in contrast to the BDD sensitivity, the CI sensitivity is not necessarily zero when the diaphragm is fully open (see Section \ref{43}). The experimental results show a similar trend for QWP orientations of $\theta_{1/4}$=25°, 45° and BDD, but a different one when $\theta_{1/4}$=0°. The sensitivity is even negative for $\theta_{1/4}$=0° (see Section~\ref{43} for explanation). Similar to the BDD case, the model thus correctly predicts the sensitivity in CI for the thermal peak amplitude and the peak-to-peak amplitude of the first acoustic echo. 

Although the model correctly predicts the the sensitivity in BDD and CI, the extracted fit parameters $A$ and $B$ differ between the different measurements. We attribute this to experimental uncertainties; slightly different alignment for each measurement, the reflection of pump power ($\sim 10\%$) on the sapphire plate and variations in experimental conditions (e.g. room temperature). However, the ratio between the fit parameter $A$ for the thermal peak amplitude and that of the peak-to-peak amplitude of the first acoustic echo equals around 16\% ($\pm 3\%$) except for the $\theta_{1/4}$ = 0° configuration that has a ratio of 33 \%. This also holds for the $B$ parameter for which the ratio is 50 \% ($\pm 5\%$) and 79\% for the $\theta_{1/4}$ = 0° configuration. We attribute the different ratios of the $\theta_{1/4}$ = 0° configuration to the fact that exactly this configuration was measured several days after the other configurations and therefore had to be re-aligned significantly. The further constant ratio of $A$ and $B$ further support the validity of the model.

\begin{figure}
\includegraphics[scale=1]{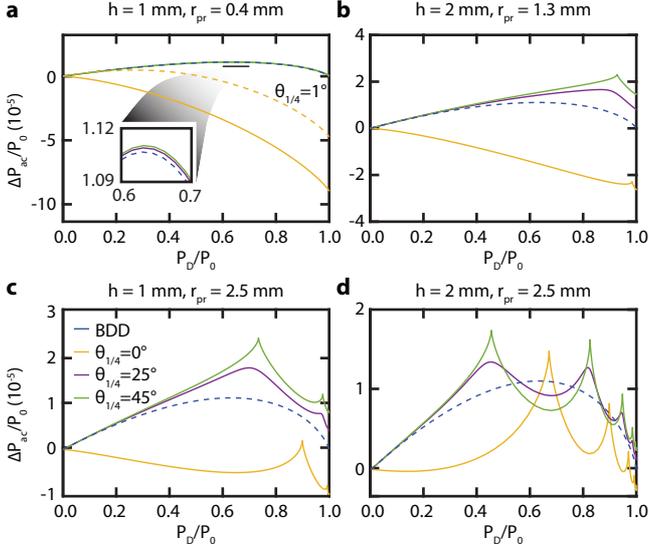}
\centering
\caption{Calculated CI sensitivity for $\theta_{1/4}$ = 0° (yellow), 25° (purple), 45° (green) and for different values of $r_{pr}$ and $h$: {\bf a} $h = 1$ mm and $r_{pr} = 0.4$ mm, {\bf b} $h = 2$ mm and $r_{pr} = 1.3$ mm, {\bf c} $h = 1$ mm and $r_{pr} = 2.5$ mm, and {\bf d} $h = 2$ mm and $r_{pr} = 2.5$ mm. For comparison, the sensitivity curve of BDD (blue) is shown in all panels. Panel {\bf a} also shows the CI sensitivity for $\theta_{1/4}$ = 1° (dashed yellow) to indicate the large change in sensitivity for a small change of $\theta_{1/4}$ around 0°. The overlapping lines in the inset of panel {\bf a} indicates that the sensitivity barely depends on $\theta_{1/4}$ between 25° and 45°.}
\label{fig:fig6}
\end{figure}

\begin{figure}
\includegraphics[scale=1]{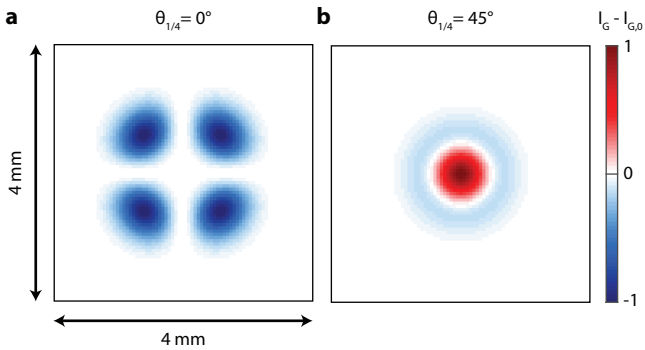}
\centering
\caption{Calculation of the difference ($I_G-I_{G,0}$) between the pattern with and without an acoustic wave for $h = $ 1 mm, $r_{pr} = $ 0.4 mm, and {\bf a} $\theta_{1/4} = $ 0°, and {\bf b} $\theta_{1/4} = $ 45°. Panels are normalized in intensity.}
\label{fig:fig7n}
\end{figure}

\subsection{Optimizing the sensitivity}
\label{43}

To optimize the CI sensitivity, we calculate the relative power variation $\Delta P_{ac}/P_0$ for different values of $h$ and $r_{pr}$ and $\theta_{1/4}$=0°, 25°, and 45° as a function of the diaphragm opening (see Figure~\ref{fig:fig6}). The CIPs corresponding to these sensitivities are depicted in Figure~\ref{fig:fig3}. In Figure~\ref{fig:fig6}, we observe the following features. The values for $\Delta P_{ac}/P_0$ are similar to those obtained experimentally (see Figure~\ref{fig:fig5}) despite we do not know the exact values of $z_0$ and $A_0$ (see Equation~\ref{eq5}) in the experiment. In contrast to BDD, we observe a nonzero sensitivity for CI in case of a completely opened diaphragm in several configurations. By increasing the probe radius $r_{pr}$ and/or the sapphire plate thickness $h$, the sensitivity changes from the one of BDD into one with distinct maxima in sensitivity even exceeding that of BDD for $\theta_{1/4} = $ 25° and 45°. For $\theta_{1/4} = $ 0°, the sensitivity changes sign and becomes positive. For higher values of $r_{pr}$ and/or $h$, even more local maxima in the sensitivity appear.

When comparing the calculated sensitivities in Figure \ref{fig:fig6}a to the corresponding experimental results presented in Figure \ref{fig:fig5}, we find that despite the agreement in trend, the relative amplitudes are different. The sensitivity of BDD and CI for $\theta_{1/4} = $ 25°, 45° should in theory almost overlap, while differences are measured experimentally. We attribute this to the variation in alignment and experimental conditions, as discussed in Section \ref{42}. For $\theta_{1/4} = $ 0°, the theoretical difference in sensitivity with the other QWP angles is much higher than the one measured in reality. We attribute this to the unknown values of $z_0$ and $A_0$ in the experiment, the several days delay between the $\theta_{1/4} = $ 0° measurement and the other ones, and also to the experimental error in QWP angle. By comparing the continuous ($\theta_{1/4} = $ 0°) and dashed ($\theta_{1/4} = $ 1°) yellow lines in Figure~\ref{fig:fig6}a, we find that the sensitivity strongly depends on the QWP angle, at $\theta_{1/4} = $ 1° already almost halves the sensitivity. As an experimental error of 1° or less in the QWP orientation is realistic, we attribute the difference in sensitivity between experiment and calculations at $\theta_{1/4} = $ 0° to it. 

\begin{figure}
\includegraphics[scale=1]{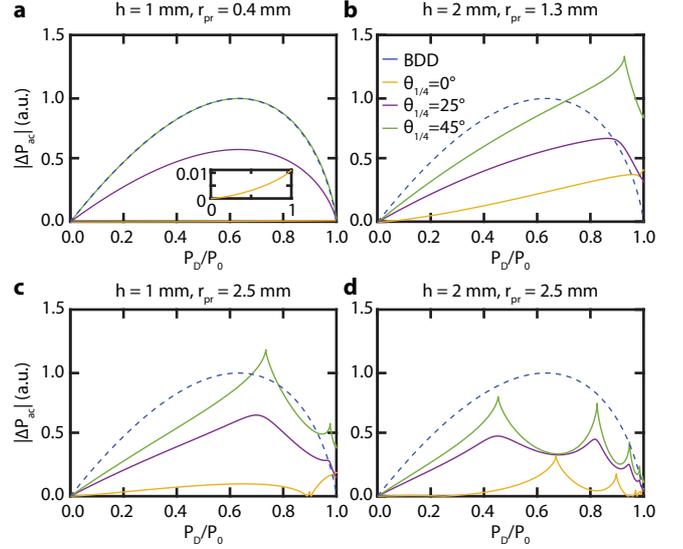}
\centering
\caption{Calculated absolute CI sensitivity $|P_{ac}|$ for $\theta_{1/4}$ = 0° (yellow), 25° (purple), 45° (green) and for different values of $r_{pr}$ and $h$: {\bf a} $h = 1$ mm and $r_{pr} = 0.4$ mm, {\bf b} $h = 2$ mm and $r_{pr} = 1.3$ mm, {\bf c} $h = 1$ mm and $r_{pr} = 2.5$ mm, and {\bf d} $h = 2$ mm and $r_{pr} = 2.5$ mm. The intensities are normalized w.r.t. the maximum of the BDD sensitivity shown by dashed blue line. The inset in panel {\bf a} shows a zoom of the $\theta_{1/4}$ = 0° case.}
\label{fig:figB}
\end{figure}

\begin{figure}[ht]
\includegraphics[scale=1]{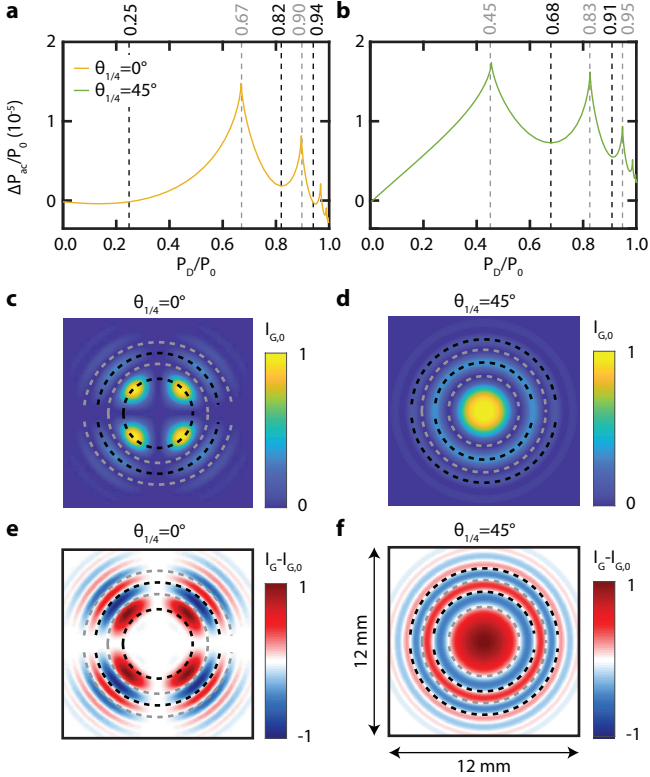}
\centering
\caption{Periodicity in the CI sensitivity at $h$=2 mm, $r_{pr}$=2.5 mm. {\bf a} and {\bf b} show the CI sensitivity for $\theta_{1/4}$=0° and $\theta_{1/4}$=45°, respectively. The dashed black (gray) lines indicate the minima (maxima) of sensitivity for a given diaphragm opening quantified by $P_D/P_0$. {\bf c} and {\bf d} show the corresponding conoscopic interference patterns $I_{G,0}$ and the diaphragm openings corresponding to minima (red) and maxima (green) in CI sensitivity. {\bf e} and {\bf f} calculations of the difference ($I_G-I_{G,0}$) between the patterns with and without an acoustic wave corresponding to the CIPs shown in panel {\bf c} and {\bf d}. Panels {\bf c}-{\bf f} are normalized in intensity.}
\label{fig:fig8}
\end{figure}

Let us now focus on understanding the negative sign of the sensitivity for $\theta_{1/4}$=0°. We attribute this to the change in CIP. For $\theta_{1/4}$=0° (see Figure \ref{fig:fig3}), we see that the center of the CIP is an isogyre. In contrast, the CIPs for $\theta_{1/4}$=25° and $\theta_{1/4}$=45° have a maximum in intensity at the center and therefore show similar sensitivities in Figure~\ref{fig:fig6}. The fast thermal expansion and acoustic pulses cause a small change in divergence angle of the probe beam. Consequently, $r_{pr}$ decreases by the factor $1+\xi$ (see Equation~\ref{eq5}) as $\xi$ is negative. In turn, this slightly shrinks the CIP. Therefore, relatively more light will pass closer to the optical axis through the sapphire plate. This light acquires a smaller phase shift $\delta_S$ than rays further away from the optical axis. For $\theta_{1/4}$=0°, all light is P-polarized before going through the sapphire plate. Due to the PBS, the photodetector only detects light that has a S-polarization component which is thus less in presence of the acoustic pulse. In contrast, the total probe power increases for $\theta_{1/4}$=25° and $\theta_{1/4}$=45°. The light has both P- and S-polarization components before going through the sapphire plate. The S-polarization component is also focused on the center and experience less phase shift $\delta_S$. Hence, more of this S-polarized light will arrive at the photodetector resulting in an increase of the total measured probe power. Considering this argument, the relative probe power $\Delta P_{ac}/P_0$ (see Equation~\ref{eq8}) will be negative for $\theta_{1/4}$=0° and positive for $\theta_{1/4}$=25° and $\theta_{1/4}$=45°, as shown in Figure \ref{fig:fig7n}. The observed negative sensitivity for $\theta_{1/4}$=0° reverses the sign of the acoustic signal. In case where this signal has both a reflectometry and BDD/CI component, as seen in Section \ref{42}, this can reduce the total sensitivity of detection. However, this can be circumvented by changing the sign of $z_0$ (see Equation \ref{eq5} and Ref. \cite{Chigarev2006}) by moving the sample to the other side of the probe beam focus.

The reason causing the negative sensitivity for $\theta_{1/4}$=0° also makes CI sensitive to acoustic waves without a diaphragm (see Figure \ref{fig:fig6}), i.e. the sensitivity is not zero when the diaphragm is fully open ($P_D/P_0$ = 1). Due to the slight variation in $r_{pr}$, the reflected probe beam experiences a slightly different phase shift $\delta_S$ when propagating through the sapphire plate. In turn, this results in a slightly different CIP (see Figure \ref{fig:fig7n}). Therefore, the incident intensity on the photodetector is varying, even without the use of a diaphragm.

The reduction in probe power $P_0$ incident on the photodetector for $\theta_{1/4}$=0° (see Figure~\ref{fig:fig3}e) also has an effect on the relative probe power $\Delta P_{ac}/P_0$. As seen in Figure \ref{fig:fig6}, the sensitivity for $\theta_{1/4}$ = 0° is much larger than the one for BDD and the other values of $\theta_{1/4}$ for small $h$ and $r_{pr}$. When $P_0$ is not used to normalize $\Delta P_{ac}$, the sensitivity of BDD will exceed that of CI in almost all studied configurations (see Figure~\ref{fig:figB}). The high sensitivity of CI is thus a direct consequence of the normalization by $P_0$ in the calculation of the relative probe power $\Delta P_{ac}/P_0$.

In order to understand the local maxima in the sensitivity shown in Figure \ref{fig:fig6}, we compare the CIPs for $\theta_{1/4}$=0° and $\theta_{1/4}$=45° obtained with different diaphragm apertures in Figure \ref{fig:fig8}. All maxima in the sensitivity correspond to an aperture with the edges of the diaphragm placed in the isogyres (dark fringes). The minima exactly occur when the diaphragm edges are on top of the bright fringes (isochromates). To find out why the sensitivity is maximum (minimum) at the isogyres (isochromates), we show the intensity difference $I_{G}-I_{G,0}$ between a pattern with and without acoustic wave (see Equation \ref{eq8}) in Figure \ref{fig:fig8}e and \ref{fig:fig8}f. The acoustic waves induce a variation in the CIP due to the slight change in divergence angle as well as a change in $\delta_S$. The sign of this intensity variation is alternatingly positive and negative. By placing the diaphragm edges on the isochromates of the CIP, the same number of positive and negative variations are incident on the photodetector. As we integrate this CIP over the open area of the diaphragm, the light intensity variation partly cancels out and thus results in a minimal sensitivity. In contrast, by placing the diaphragm edges on the dark fringes of the CIP, more positive than negative variations of intensity are obtained, resulting in a maximum sensitivity to acoustic waves.

Finally, we compare the sensitivity of CI with that of BDD. As Figure~\ref{fig:fig6} shows, CI is not always more sensitive than BDD. However, by choosing the right diaphragm opening and QWP orientation, CI can be made more sensitive than BDD. For example, the CI configuration with $r_{pr} = $ 2.5 mm, $h = $ 1 mm, $\theta_{1/4} = $ 45°, and $P_D/P_0 = \sim 0.73$, has a total sensitivity almost twice that of BDD. The total sensitivity of CI configuration with $r_{pr} = $ 0.4 mm, $h = $ 1 mm, $\theta_{1/4} = $ 0°, and no diaphragm, is even up to 8 times higher than that of BDD.

\section{Conclusion}

Conoscopic Interferometry (CI) is a promising detection technique for ultrafast acoustics that can offer an improved SNR compared to Beam Distortion Detection (BDD) and reflectometry. We developed a model that predicts the sensitivity of CI and BDD. Our results show that for a given probe power incident on the photodetector, CI can be more sensitive than BDD for detecting the surface displacement of a sample, if one carefully chooses the right parameters. By using a 1 mm thick sapphire plate, a probe beam radius of 0.4 mm, a Quarter Wave Plate orientation of 0° and no diaphragm, CI is up to 8 times more sensitive than BDD. Moreover, we showed that the CI sensitivity is optimal when the diaphragm aperture cuts the radially symmetric conoscopic interference patterns in its dark fringes. We validated these observations experimentally on a 2.4 \textmu m thick silicon substrate coated with $30$ nm aluminum. We foresee significant improvements of the CI detection sensitivity by using different birefringent crystals, by beam shaping the probe beam, or using different diaphragm geometries. Because of the enhanced sensitivity compared to BDD and reflectometry on materials with low photoelastic constants, optimized CI detection schemes could play a central role in the future of ultrafast acoustics.

\section*{Acknowledgment}
M.R., R.G., G.J.V. acknowledge support from project TKI-HTSM/19.0172.

\section*{Author contribution}
\textbf{M.R.}: Conceptualization, Methodology, Software, Formal Analysis, Investigation, Writing - original draft. 
\textbf{R.G.}: Conceptualization, Investigation, Writing - review \& editing.
\textbf{U.A.}: Conceptualization, Writing - review \& editing, Supervision, Project administration.
\textbf{Z.Z.}: Conceptualization, Supervision. 
\textbf{N.P.}: Conceptualization, Supervision.
\textbf{G.J.V.}: Conceptualization, Methodology, Writing - review \& editing, Supervision, Project administration, Funding acquisition.

\FloatBarrier

% \bibliographystyle{elsarticle-num} 
% \bibliography{cas-refs}

\section*{Vitae}

\begin{wrapfigure}{l}{0.17\textwidth}
    \includegraphics{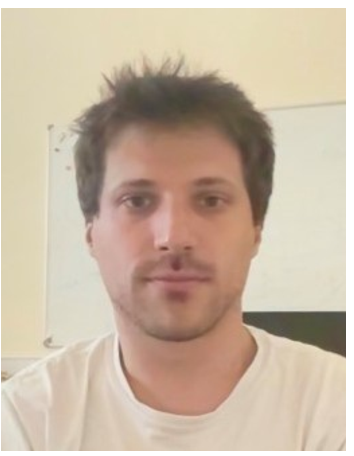}
\end{wrapfigure}

\noindent \textbf{Martin Robin} received his PhD degree in the field of Laser-Ultrasonics at the Université Polytechnique des Hauts-de-France, France, 2019, after obtaining his M.Sc. degree in acoustics at Le Mans Université, France, 2015. His doctorate work was about the characterization of the adhesion in film-on-substrate using Surface Acoustic Wave, generated and detected by lasers. He then pursued as a postdoctoral researcher at the Technische Universiteit Delft, Netherlands, between 2020 and 2022. There, he worked on the design and installation of a Picosecond Ultrasonic setup for imaging and Non-Destructive Testing applications. He is now employed as a postdoctoral researcher at Institut Lumière Matière, Université Claude Bernard Lyon 1, France, to characterize bio-metamaterials by Surface Acoustic Waves using Laser-Ultrasonics.

\ \\

\begin{wrapfigure}{l}{0.17\textwidth}
    \includegraphics{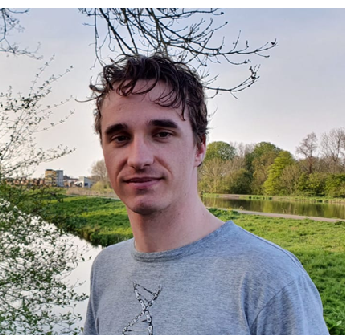}
\end{wrapfigure}

\noindent \textbf{Ruben Guis} obtained his master's degree in Physics at Leiden University, Netherlands in 2020. For his thesis project he worked towards 3D imaging at the nanoscale by attempting to measure the subsurface information using resonant hydrogen spins in Magnetic Resonance Force Microscopy. Started in October 2020, he is currently doing his PhD at the Technische Universiteit Delft, Netherlands, where he is again working on 3D imaging at the nanoscale, this time using ultrasound to get the subsurface information, by integrating nanoscale acoustics into atomic force microscopy.

\ \\

\begin{wrapfigure}{l}{0.17\textwidth}
    \includegraphics{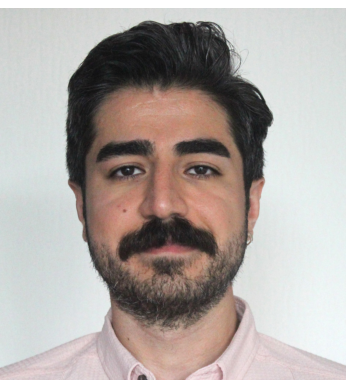}
\end{wrapfigure}

\noindent \textbf{M. \"Umit Arabul} received the B.Sc. degree (2011) in Electrical \& Electronics Engineering from the Middle East Technical University (METU), Ankara, Turkey. Next, he joined the Medical Imaging Lab. in the Institute of Biomedical Engineering, Bogaziçi University (BOUN), Istanbul and obtained the M.Sc. degree in Biomedical Engineering (2013). He pursued his Ph.D. in the Photoacoustics and Ultrasound Laboratory of Eindhoven (PULS/e Lab), in the Cardiovascular Biomechanics group of the department of Biomedical Engineering in the Eindhoven University of Technology, The Netherlands. His research focused on photoacoustic imaging of carotid arteries and fundamental characterization and preclinical validation of photoacoustic imaging (2018). Currently, he is working as a Research Scientist at ASML with the focus on semiconductor metrology.

\ \\

\begin{wrapfigure}{l}{0.17\textwidth}
    \includegraphics{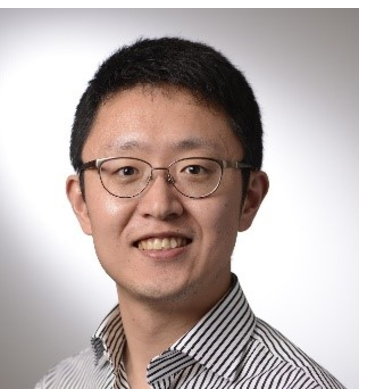}
\end{wrapfigure}

\noindent \textbf{Zili Zhou} obtained his Master’s degree of Optical Engineering from Chinese Academy of Sciences, Shanghai, China, in 2010. Then he received his PhD degree from Applied Physics department of Eindhoven University of Technology, the Netherlands, in 2014, where he focused on multi-photon detection with superconducting nanowires. Since 2015, he works at Research department of ASML on optical nanometrology projects, as a senior research scientist.

\ \\

\begin{wrapfigure}{l}{0.17\textwidth}
    \includegraphics{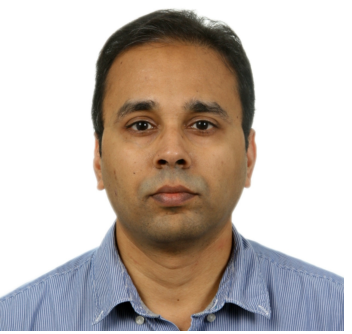}
\end{wrapfigure}

\noindent \textbf{Nitesh Pandey} (PhD 2011, NUI Maynooth) is a Principal Research Engineer working at the Advanced Technology Development group at ASML. He joined ASML in 2011 and has worked on various topics related to Optical sensors in Lithography scanners and Overlay metrology. He is currently working in the area of pattern fidelity control, OPC and stochastic placement error modelling.

\ \\

\begin{wrapfigure}{l}{0.17\textwidth}
    \includegraphics{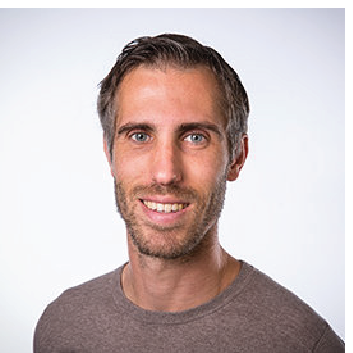}
\end{wrapfigure}

\noindent \textbf{Gerard J. Verbiest} obtained his M.Sc. degree in theoretical physics (cum laude, 2009), and a PhD in experimental physics (2013) from Leiden University. His doctorate work focused on the application of ultrasound at MHz frequencies in atomic force microscopes to enable subsurface atomic force microscopy. He then worked as postdoctoral researcher at the RWTH Aachen in Germany between 2013 and 2018 on graphene mechanics and dynamics. Since August 2018, Gerard works as assistant professor at the Delft University of Technology. His work focusses on nanoscale acoustics with applications in atomic force microscopes, graphene and other two-dimensional materials, and plant physiology.
\end{document}